\documentstyle[12pt]{article}
\def\setb@se#1{\baselineskip=#1 \normalbaselineskip=#1}
\setb@se{14pt}
\newcommand{\be}{\begin{equation}}
\newcommand{\ee}{\end{equation}}
\newcommand{\ii}{{\it i}}

\newcommand{\iii}{\int_{0}^{\infty}}
\newcommand{\mm}{\frac{M}{\sqrt{2}}}
\begin{document}
\begin{titlepage}
\begin{flushright}
Z\"urich University Preprint

ZU-TH 13/94,~~ May 1994

hep-th/9406017
\end{flushright}
\vspace{20 mm}
\begin{center}
\Huge 

Einstein-Yang-Mills sphalerons and level crossing 

\vspace{5 mm}
\end{center}
\begin{center}
{\bf  Mikhail S. Volkov}\footnote{On  leave  from  Physical-Technical 
Institute of the  Academy  of  Sciences  of  Russia,  Kazan  420029, 
Russia}

\vspace{5 mm}
Institut    f\"ur    Theoretische    Physik    der    Universit\"at 
Z\"urich-Irchel, 

Winterthurerstrasse 190, CH-8057 Z\"urich,

Switzerland

e-mail:\ volkov@physik.unizh.ch
\end{center}
\vspace{20 mm}\begin{center}{\large Abstract}\end{center}
\vspace{10mm}

The fermion energy spectrum along paths which connect  topologically 
distinct vacua in the Einstein-Yang-Mills theory passing through the 
gravitational sphaleron equilibrium solutions is investigated. 

\end{titlepage}
\newpage
\section {Introduction}

As is known, fermion number may be  non-conserved  in  the  standard 
model at high energies, in non-perturbative  processes  mediated  by 
sphalerons  \cite{MK}.  Such  processes  correspond  to  overbarrier 
transitions between topologically distinct  vacuum  sectors  of  the 
theory, where the barrier height is given by the sphaleron mass.  As 
the Chern-Simons (CS) number of the gauge field varies, this  causes 
the anomalous nonconservation of fermion number. 

In the first quantization picture, the change of fermion number is a 
result  of  level  crossing  \cite{Callan}.  Consider  an  adiabatic 
process, in which the  boson  fields  interpolate  between  distinct 
vacuum sectors leading to a change of the CS  number  $N_{CS}$.  The 
energy spectrum of massless  fermions  interacting  with  the  boson 
background then depends parametrically on  $N_{CS}$.  When  $N_{CS}$ 
varies, some of the energy levels  cross  zero  ending  up  with  an 
opposite sign of their energy. If the initial state  corresponds  to 
the filled Dirac sea, then (anti) fermions will be  present  in  the 
final state. For the  processes  involving  electroweak  sphalerons, 
this  picture  was  studied  in  Refs.\cite{cross},\cite{diak}.   In 
addition, fermion production in dynamical decay of the sphaleron has 
been considered \cite{cott}. 

The aim of the present paper is to investigate  the  level  crossing 
phenomenon arising within the  context  of  the  Einstein-Yang-Mills 
(EYM) theory. As is  known,  in  this  theory  there  exist  regular 
particle-like solutions \cite{BK}.  Such  solutions  relate  to  the 
critical points  on  the  top  of  the  potential  barrier  dividing 
distinct  vacua  in  the  theory,  which  allows  them  also  to  be 
interpreted as sphalerons \cite{10}. One may expect that  transition 
processes mediated by these EYM  sphalerons  become  significant  at 
high energies \cite{main}. 

It has been shown recently that  the  EYM  sphalerons  possess  zero 
energy fermion bound state \cite{GS}. The existence  of  this  state 
indicates that some of the fermion energy levels cross zero when the 
evolving  background  field  passes  through  the  sphaleron   field 
configuration. In the present paper we investigate the  whole  level 
crossing  picture for  the  case  when  the  background  EYM  fields 
interpolate between distinct vacua. 

We define below paths which connect topologically distinct vacua  in 
the  EYM  configuration  space   passing   through   the   sphaleron 
equilibrium solutions. The field evolving adiabatically along such a 
path is used as the background  field  in  the  Dirac  equation.  To 
discretize the fermion energy spectrum, we compactify the 3-space by 
lifting the system onto a 3-sphere  with  large  but  finite  radius 
$R_{\ast}$, and thus the distance between energy levels  is  of  the 
order of $1/R_{\ast}$. For this geometrical regularization, we
determine analytically the free fermion energy levels 
and zero fermion  modes as well. In  addition,  we 
derive a simple analytic approximation for the whole level  crossing 
spectrum, which gives qualitatively the  right  result.  Finally,  we 
solve the Dirac equation numerically and  find  the  spectrum precisely. 
Our result is depicted in Fig.3 and shows explicitly the  phenomenon  of 
level crossing appearing in the EYM theory.

\section{EYM sphalerons and vacuum to vacuum paths}

Consider the action of the EYM theory for the $SU(2)$ gauge group
\be
S_{EYM}=-{1\over 16\pi G}\int R\sqrt{-g}d^{4}x - 
{1\over 2}\int trF_{\mu\nu}F^{\mu\nu}\sqrt{-g}d^{4}x.
                                                     \label{1}
\ee
Here  $F_{\mu\nu}=\partial_{\mu}A_{\nu}-\partial_{\nu}A_{\mu}-   \ii 
e[A_{\mu},A_{\nu}]$ is the matrix valued gauge  field   tensor,  $e$ 
is the gauge   coupling  constant,  $A_{\mu}=A_{\mu}^{a}\tau^{a}/2$, 
and  $\tau^{a}\   (a=1,2,3)$   are  the  Pauli  matrices.  Our  sign 
conventions are those of Landau \& Lifshitz \cite{ll}. 

We choose the spherically symmetric gauge field as follows 
\be 
A=A_{\mu}dx^{\mu}=\frac{\ii}{2e}(1-w)U dU^{-1},\ \ \ 
U=exp(\ii\frac{\pi}{2} n^{a}\tau^{a}),                    \label{2}
\ee
where   $n^{a}=(sin\vartheta cos\varphi, sin\vartheta   sin\varphi, 
cos\vartheta)$, and the gravitational field is parameterized by 
\be
ds^{2} = l^{2}_{e}\left( (1-\frac{2m}{r})\sigma^{2}dt^{2} -
\frac{dr^{2}}{1-2m/r} - r^{2}(d\vartheta^{2} + sin^{2}\vartheta 
d\varphi^{2})\right).                                  \label{3}
\ee
Here $l_{e}=\sqrt{4\pi}l_{pl}/e$ is the only dimensional quantity in 
the problem ($l_{pl}$ being Planck's length); the functions $w$, $m$ 
and $\sigma$ depend on $t$ and $r$. 

For the static case, the corresponding Lagrangian equations read 
\be
((1-\frac{2m}{r})\sigma w')'=\sigma\frac{w(w^{2}-1)}{r^{2}}, 
                                                        \label{4} 
\ee
\be
m'=(1-\frac{2m}{r})w'^{2}+\frac{(w^{2}-1)^{2}}{2r^{2}},\ \ \ \ \ \ \ 
\sigma'=2\frac{w'^{2}}{r}\sigma.                        \label{6}
\ee
These equations are known to possess  regular,  asymptotically  flat 
solutions discovered by Bartnik and  McKinnon  \cite{BK}.  These  BK 
solutions are labeled by an integer, $n$. The  function  $w(r)$  for 
the $n$-th solution, $w_{n}(r)$, has $n$ nodes in the  domain  $0<r< 
\infty$ and satisfies the following boundary conditions:
\be
w_{n}(r)=1+O(r^{2})\  \  {\rm  as}\ \ r\rightarrow 0,
\ \ \ w_{n}(r)=(-1)^{n}+\alpha_{n}/r+O(1/r^{2})\ \ \
{\rm as}\ \ r\rightarrow\infty,                  \label{6:1}
\ee
where $\alpha_{n}$ are numerically  known  constants  \cite{Kunzle}. 
The corresponding metric functions  $m_{n}(r)$  and  $\sigma_{n}(r)$ 
increase monotonically from  $m_{n}(0)=0$  to  $m_{n}(\infty)=m_{n}$ 
and  from  $\sigma_{n}(0)=  \sigma_{n}$  to  $\sigma_{n}(\infty)=1$, 
respectively (see Fig.1). The ADM  masses  of  these  solutions  are 
$M_{n}=\sqrt{4\pi}M_{pl}m_{n}/e$, with $M_{pl}$ being Planck's mass, 
where  $m_{n}$  increases  as  $n$  grows  from  the  minimal  value 
$m_{1}=0.828$ to $m_{\infty}=1$. 

As was shown in Ref.\cite{10},  the  odd-$n$  BK  solutions  may  be 
treated as sphalerons. This interpretation  is  based  on  the  fact 
that, through any such solution, one may find a one-parameter family 
of fields which  interpolates  between  topologically  distinct  EYM 
vacua. If  $\lambda$  denotes  the  parameter  of  the  family,  the 
corresponding gauge field can be chosen as follows \cite{10}
\be 
A_{\mu}dx^{\mu}=\frac{\ii}{2e}(1-w_{n}(r))U dU^{-1},\ \ \ 
U=exp(\ii\lambda n^{a}\tau^{a}),                          \label{7}
\ee
where $\lambda$ ranges from zero to $\pi$. The metric functions  $m$ 
and $\sigma$ in (\ref{3}) are given by 
$$ 
\sigma (r)= exp\{-2sin^{2}\lambda\int_{r}^{\infty}w_{n}'^{2} 
\frac{dr}{r}\}, $$
\be
m(r)=\frac{sin^{2}\lambda}{\sigma(r)}\int_{0}^{r}(w_{n}'^{2} + 
sin^{2}\lambda\frac{(w_{n}^{2}-1)^{2}}{2r^{2}})\sigma dr. \label{8} 
\ee
It is worth noting that, for any value of $\lambda$, the  fields  of 
this family  satisfy  the  two  Einstein  equations  $G^{0}_{0}=8\pi 
GT^{0}_{0}$ and $G^{r}_{r}=8\pi  GT^{r}_{r}$;  when  $\lambda=\pi/2$ 
these equations reduce to those in (\ref{6})  \cite{10}.  Obviously, 
for $\lambda=\pi/2$, the fields  (\ref{7}),(\ref{8})  coincide  with 
the $n$-th BK solution field.  When  $\lambda$  approaches  zero  or 
$\pi$ values, the spacetime metric becomes flat and the gauge  field 
(\ref{7}) vanishes. Thus, the family of  fields  (\ref{7}),(\ref{8}) 
describes a loop (in the EYM  configuration  function  space)  which 
interpolates between the trivial vacuum and the $n$-th  BK  solution 
when $\lambda$ runs from 0 to  $\pi/2$,  and  returns  back  to  the 
vacuum as $\lambda$ changes from $\pi/2$ to $\pi$. 

It is important  that  this  loop  can  be  transformed  to  a  path 
interpolating between neighbouring  EYM vacua with different winding 
numbers of the gauge field. The gauge transformation 
\be
A_{\mu}\rightarrow U(A_{\mu}+\frac{\ii}{e}\partial_{\mu})U^{-1},\ \ 
{\rm with}\ \ 
U=exp (\ii\frac{\lambda}{2}(w_{n}-1)n^{a}\tau^{a})
                                               \label{8:1}
\ee
yields 
$$
A_{\mu}dx^{\mu}=\frac{\ii}{2e}(1-w_{n}(r))U_{+}\ dU_{+}^{-1}+
\frac{\ii}{2e}(1+w_{n}(r))U_{-}\ dU_{-}^{-1},
$$
\be
U_{\pm}=exp(\ii\lambda (w_{n}\pm 1)n^{a}\tau^{a}/2),     \label{9}
\ee
whereas  the  gauge  invariant  metric  functions  (\ref{8})  remain 
unchanged. One can see that, when $\lambda$ runs from zero to $\pi$, 
the field (\ref{9}) interpolates between two different pure  gauges. 
We will refer to the gauge field potentials (\ref{7}) and  (\ref{9}) 
as gauge field represented in the gauge $I~$ and in the gauge $II~$, 
respectively. Notice that, in the gauge $II~$, one has  $A_{a}=O(r)$ 
when     $r\rightarrow     0$     and     $A_{a}=O(1/r^{2})$      as 
$r\rightarrow\infty$, with $A_{a}$ being the  field  component  with 
respect to an orthonormal frame. 

Let $\lambda$ depend adiabatically on  time,  in  such  a  way  that 
$\lambda(t= -\infty)=0$ and $\lambda(t=\infty)=\pi$.  Then,  in  the 
gauge $I$, the time component of the field appears, whereas  in  the 
gauge $II$ one has $A_{0}=0$. Recall the Chern-Simons  (CS)  current 
and its divergence 
\be
K^{\mu}={e^{2}\over 8\pi^2}tr{\varepsilon^{\mu\nu\alpha\beta}\over 
\sqrt{-g}}A_{\nu}(\nabla_{\alpha}A_{\beta}-\frac{2\ii e}{3}
A_{\alpha}A_{\beta}),\ \ \nabla_{\alpha}K^{\alpha}=
\frac{e^{2}}{16\pi^{2}}trF_{\mu\nu}\tilde{F}^{\mu\nu},   \label{10}
\ee
where   $\nabla_{\alpha}$   is   the   covariant   derivative,   and 
$\tilde{F}^{\mu\nu}$  is  dual  tensor.  Integration  leads  to  the 
following relation:
\be
\frac{e^{2}}{16\pi^{2}}\int_{-\infty}^{t}dt\int d^{3}x \sqrt{-g}
trF_{\mu\nu}\tilde{F}^{\mu\nu}=
\left.\int d^{3}x\sqrt{-g}K^{0}\right|^{t}_{-\infty} 
+ \int_{-\infty}^{t} dt\oint\vec{K}d\vec{\Sigma}.        \label{11}
\ee
In the temporal gauge $II~$, the surface integral entering the right 
hand side vanishes, while the gauge  invariant  left  hand  side  is 
\cite{10}
\be
\frac{3}{2\pi}\int_{-\infty}^{t}dt\ \dot{\lambda}sin^{2}\lambda
\int_{0}^{\infty}dr\ w_{n}'(w_{n}^{2}-1)=\left. \frac{3}{4\pi}
(\lambda-sin\lambda cos\lambda)(\frac{1}{3}w_{n}^{3}-w_{n})
\right|^{w_{n}(\infty)=-1}_{w_{n}(0)=1}.                \label{5}
\ee
Noting that the gauge potential (\ref{9}) vanishes  at  $\lambda=0$, 
one obtains the following value of the  CS  number  along  the  path 
(\ref{9}) \cite{10}, \cite{main}:
\be
N_{CS}=\left.\int\sqrt{-g}K^{0}d^{3}x\right|_{t}=
\frac{1}{\pi}(\lambda-sin\lambda cos\lambda),            \label{12}
\ee
which changes from zero to one as $\lambda$ runs from $0$  to  $\pi$ 
(see also Ref.\cite{BS}  for  a  quite  different  approach  to  the 
calculation of the CS numbers). 

Thus the fields (\ref{3}), (\ref{8}), (\ref{9}) interpolate  between 
two EYM vacua with different winding numbers of the gauge field, and 
coincide with  the  (gauge  transformed)  field  of  the  $n$-th  BK 
solution when $\lambda=\pi/2$. This allows one to treat  odd-$n$  BK 
solutions as sphalerons, ``lying''  on  the  top  of  the  potential 
barrier  separating  distinct  topological  vacuum  sectors  of  the 
theory.  The  profile  of  this  barrier  may   be   obtained   from 
Eq.(\ref{8}) because the metric function $m(r)$  obeys  the  initial 
value  constraint,  which  allows  us  to  define  the  ADM   energy 
\cite{10}:
$$ 
U(\lambda) = \lim_{r\rightarrow\infty}m(r)=
$$
\be
=sin^{2}\lambda\iii (w_{n}'^{2} + 
sin^{2}\lambda\frac{(w_{n}^{2}-1)^{2}}{2r^{2}})
exp(-2sin^{2}\lambda\int_{r}^{\infty}w_{n}'^{2}\frac{dr}{r})dr,
                                                         \label{13} 
\ee
(a plot of the function $U(N_{CS}(\lambda))$ for $n=1$  is  depicted 
in Fig.2).

\section{Compactification of the 3-space}

Our aim is to investigate the energy spectrum of chiral fermions  in 
the external fields (\ref{7}), (\ref{3}), (\ref{8}) which  form  the 
potential barrier (\ref{13}). As one expects, the change of  the  CS 
number across the barrier causes a non-trivial spectral  flow.  This 
concept is, however, not meaningful since  the  energy  spectrum  is 
purely continuous. Therefore, we apply some regularization procedure 
to discretize the spectrum. The most natural way  is  to  compactify 
the 3-space by lifting the system onto a  3-sphere  with  large  but 
finite  radius  $R_{\ast}$;  the  limit  $R_{\ast}\rightarrow\infty$ 
being eventually implied (see also Ref.\cite{diak} for other  choice 
of the regularization scheme). 

For this we multiply the spatial part of the line element  (\ref{3}) 
by an appropriate conformal factor 
\be
ds^{2} = l^{2}_{e}\left( (1-\frac{2m}{r})\sigma^{2}dt^{2} -
\left( 1+\frac{r^{2}}{4R_{\ast}^{2}}\right)^{-2}\{
\frac{dr^{2}}{1-2m/r} + r^{2}(d\vartheta^{2} + sin^{2}\vartheta 
d\varphi^{2})\}\right),                           \label{13:1}
\ee
with functions $m$ and $\sigma$ still given by  Eqs.(\ref{8}).  When 
parameter $\lambda$ in (\ref{8}) interpolates between zero and $\pi$ 
values, this line element describes an everywhere  regular  geometry 
on a manifold whose spatial section is topologically  $S^{3}$.  When 
$\lambda=0,\pi$, one has  $m=0,\  \sigma=1$,  so  that  the  spatial 
section is the 3-sphere with radius $R_{\ast}$ and standard metric. 

The lift of the gauge field is locally well defined by Eqs.(\ref{7}) 
or (\ref{9}). However, in the gauge $I~$, the field is  singular  at 
$\tilde{r}=0$, where $\tilde{r}=1/r$ (we will call this point  north 
pole). Consider the gauge $II~$. In the vicinity of the north  pole, 
the components of the field (\ref{9}) with respect to an orthonormal 
frame are 
\be
A_{a}=\frac{\alpha_{n}}{R_{\ast}^{2}}~ 
a_{a}(\vartheta,\varphi) +O(\tilde{r}),  \label{13:2}
\ee
with  $a_{a}$ being some functions of  the  angular  variables,  the 
constant $\alpha_{n}$ is the same as that  entering  Eq.(\ref{6:1}). 
One  can  see  that  this  field  is   not   defined   uniquely   at 
$\tilde{r}=0$. One can not improve the situation just by  performing 
a further gauge transformation in the vicinity of the pole  (because 
the maximal decay rate of the field (\ref{9}), $A_{a}\sim  1/r^{2}$, 
is fixed by the asymptotical behaviour of  the  corresponding  gauge 
invariant energy density: $T^{0}_{0}\sim 1/r^{6}$). Notice, however, 
that for our  geometrical  regularization,  the  normalized  fermion 
modes always vanish at the north pole (see Eq.(\ref{33:2})), so that 
this ambiguity does not enter the Dirac equation. 

One can understand this also as follows.  Let  us  introduce  for  a 
moment an additional regularization,  demanding  for  the  amplitude 
$w_{n}$, in the vicinity of the north pole, to tend  faster  to  its 
asymptotical value:
\be
w_{n}(r)=(-1)^{n}+\alpha_{n}/r^{1+\epsilon}+O(1/r^{2}),\ \ \ \ \
{\rm where}\ \ \epsilon\rightarrow +0.                 \label{13:3}
\ee
The gauge field  is  then  everywhere  regular,  whereas  the  field 
topology, that is, the CS number, remains the  same.  It  turns  out 
that, in the vicinity of the pole, the regular power series solution 
for fermions converges uniformly as $\epsilon\rightarrow 0$ to  that 
corresponding  to  $\epsilon=0$.  Thus  fermions  do  not  feel  the 
$\epsilon$-dependence of the field. 

We assume therefore that Eq.(\ref{9}) specifies  the  regular  gauge 
field on a compact manifold whose geometry is described by the  line 
element (\ref{13:1}).

\section{Fermions on an EYM background}

Our next task is to perform separation of  variables  in  the  Dirac 
equation. For the sake of  generality,  we  will  consider  in  this 
section massive doublet fermions  interacting  with  arbitrary  time 
dependent,  spherically  symmetric  EYM  background   fields.   Some 
particular cases of  this  problem  were  considered  previously  in 
Refs.\cite{GE}, \cite{GS}. 

Consider the action of an $SU(2)$ doublet  of  fermions  in  an  EYM 
background 
\be
S_{f}=\int \bar{\Psi}\left(\gamma^{\mu}(\ii\nabla_{\mu}+ e
A_{\mu})-M\right)\Psi \sqrt{-g}d^{4}x, 
                                                      \label{14}
\ee
with $M$ being fermion mass; this action is  invariant  under  the 
gauge transformations 
\be
A_{\mu}\rightarrow U(A_{\mu}+\frac{\ii}{e}\partial_{\mu})U^{-1},\ \ 
\Psi\rightarrow U\Psi,\ \ \ \ U\in SU(2).          \label{14:1}
\ee

It is most economic for our purposes to make   use  of  
the  Newman-Penrose  (NP) 
formalism of spin coefficients \cite{NP}. This formalism is known to 
be a convenient tool for studying the wave equations  of  fields  of 
various spin in the algebraically distinguished  spaces  of  general 
relativity \cite{chandr}, \cite{GE}. In  the  spherically  symmetric 
case, the role of the spinor angular functions is played then by the 
spin-weighted spherical harmonics. (Notice that, in  our  problem,  the 
separation of variables in terms of spherical spinors  can  also  be 
performed \cite{sachs}, but the calculations are somewhat tedious.) 

In spinor representation, bispinor $\Psi$ decomposes as $\Psi=\left( 
\begin{array}{c} \phi^{A} \\ \bar{\xi}_{A'} \end{array}\right)$.  In 
the   chiral   limit,   two   component   spinors   $\phi^{A}$   and 
$\bar{\xi}_{A'}$  (they   carry   also   isospin   indices)   become 
right-handed and left-handed parts of bispinor $\Psi$  respectively. 
The Dirac equation reads 
\be
(\nabla_{AA'}-\ii eA_{AA'})\phi^{A}=-\ii\mm\ \bar{\xi}_{A'} ,\ \ \ \
(\nabla^{AA'}-\ii eA^{AA'})\bar{\xi}_{A'}=-\ii\mm\ \phi^{A},
                                                 \label{15}
\ee
where the spin-tensors $\nabla_{AA'}$ and $A_{AA'}$ are  related  to 
the covariant derivative operator $\nabla_{\mu}$ and the gauge field 
potential $A_{\mu}$,  via  van  der  Waerden  connection  quantities 
$\sigma^{\mu}_{AA'}$ \cite{NP}: 
$$
\nabla_{\mu}\leftrightarrow\nabla_{AA'}= 
\sigma^{\mu}_{AA'}\nabla_{\mu},\ \ \ \ 
A_{\mu}\leftrightarrow A_{AA'}=\sigma^{\mu}_{AA'}A_{\mu}.
$$

The conserved current is 
$$ 
J^{\mu}\leftrightarrow\phi_{1}^{A}\bar{\phi}_{2}^{A'}+ 
\xi_{1}^{A}\bar{\xi}_{2}^{A'}, 
$$
which gives the inner product 
\be
<\Psi_{2}|\Psi_{1}>=\int J^{0}\sqrt{-g}d^{3}x=
\int\sigma^{0}_{AA'}(\phi_{1}^{A}\bar{\phi}_{2}^{A'}
+\xi_{1}^{A}\bar{\xi}_{2}^{A'})\sqrt{-g}d^{3}x. 
                                                     \label{inn}
\ee

Introduce  a  spinor  dyad 
$(o^{A},\iota^{A})$, $o_{A}\iota^{A}=1$, which  defines  a  null  NP 
tetrad $(l,n,m,\bar{m})$ according to the scheme
$$
l^{\mu}\leftrightarrow o^{A}\bar{o}^{A'},\ \
n^{\mu}\leftrightarrow \iota^{A}\bar{\iota}^{A'},\ \
m^{\mu}\leftrightarrow o^{A}\bar{\iota}^{A'},\ \
\bar{m}^{\mu}\leftrightarrow \iota^{A}\bar{o}^{A'},
$$
non-zero scalar products being 
$l^{\mu}n_{\mu}=-m^{\mu}\bar{m}_{\mu}=1$.
Expanding $\phi^{A}$ and $\bar{\xi}_{A'}$ over the spinor  basis,
$$
\phi^{A}=\Phi_{0}~o^{A}+ \Phi_{1}~\iota^{A}, \ \ \
\bar{\xi}_{A'}=\Xi_{0}~\bar{o}_{A'}+ 
\Xi_{1}~\bar{\iota}_{A'}, 
$$
and using the NP quantities 
$$
\bar{o}^{A'}\nabla_{AA'}o^{A}=\varepsilon-\rho,\ \
\bar{o}^{A'}\nabla_{AA'}\iota^{A}=\pi-\alpha,\ \
\bar{\iota}^{A'}\nabla_{AA'}o^{A}=\beta-\tau,\ \
\bar{\iota}^{A'}\nabla_{AA'}\iota^{A}=\mu-\gamma,
$$
one obtains the Dirac equation as
$$
(D+\varepsilon-\rho-\ii eA_{l})~\Phi_{0}+
(\bar{\delta}+\pi-\alpha-\ii eA_{\bar{m}})~\Phi_{1}=
\ii\mm~\Xi_{1},
$$
$$
(\delta+\beta-\tau-\ii eA_{m})~\Phi_{0}+
(\Delta +\mu-\gamma-\ii eA_{n})~\Phi_{1}=
-\ii\mm~\Xi_{0}, 
$$
$$
(D+\varepsilon^{\ast}-\rho^{\ast}-\ii eA_{l})~\Xi_{0}+
(\delta+\pi^{\ast}-\alpha^{\ast}-\ii eA_{m})~\Xi_{1}=
-\ii\mm~\Phi_{1},
$$
\be
(\bar{\delta}+\beta^{\ast}-\tau^{\ast}-\ii eA_{\bar{m}})~
\Xi_{0}+(\Delta +\mu^{\ast}-\gamma^{\ast}-\ii eA_{n})~\Xi_{1}=
\ii\mm~\Phi_{0}.                                   \label{16}
\ee
Here the standard definitions used are: $D=l^{\mu}\partial_{\mu}$, 
$\Delta=n^{\mu}\partial_{\mu}$, $\delta=m^{\mu}\partial_{\mu}$,  and 
also         $A_{l}=l^{\mu}A_{\mu}$,         $A_{n}=n^{\mu}A_{\mu}$, 
$A_{m}=m^{\mu}A_{\mu}$,   $A_{\bar{m}}    =\bar{m}^{\mu}    A_{\mu}= 
A_{m}^{\dagger}$; asterisk denotes complex conjugation. 

Let us choose the spacetime metric in the form 
\be
ds^{2} =e^{2a}dt^{2} -e^{2b}dr^{2} - R^{2}(d\vartheta^{2} + 
sin^{2}\vartheta d\varphi^{2}),                           \label{17}
\ee
where functions $a$, $b$ and $R$ depend on $t$ and $r$. The  general 
spherical gauge field \cite{W} can be represented as follows:
\be 
eA = W_{0}\hat{L}_{1}\ dt + W_{1}\hat{L}_{1}\ dr 
+\{p_{2}\ \hat{L}_{2} - (1-p_{1})\ \hat{L}_{3}\}\ d\vartheta 
+\{(1-p_{1})\ \hat{L}_{2} + p_{2}\ 
\hat{L}_{3}\}sin\vartheta\ d\varphi, 
                                                        \label{18}
\ee 
where
\be
\hat{L}_{1} = \frac{1}{2}n^{a}\tau^{a}=
sin\vartheta cos\varphi\ \hat{T}_{1} 
+sin\vartheta sin\varphi\ \hat{T}_{2} +cos\vartheta\ \hat{T}_{3},\ 
\hat{L}_{2}=\partial_{\vartheta}\hat{L}_{1},\ \  
\hat{L}_{3}=\frac{1}{sin\vartheta}\partial_{\varphi}\hat{L}_{1},     
\label{19}
\ee
with   $\hat{T}_{a}=\tau^{a}/2$;   $W_{0},W_{1},p_{1},p_{2}$   being 
functions of $t$ and  $r$.  The  gauge  transformation  (\ref{14:1}) 
generated by 
\be
U=exp(\ii\Omega(t,r)\hat{L}_{1})                  \label{18:1}
\ee
preserves the form of the field (\ref{18}), altering the functions 
$W_{0}$, $W_{1}$, $p_{1}$, $p_{2}$ as
\be
W_{0}\rightarrow W_{0}+\dot{\Omega},\ \ \ 
W_{1}\rightarrow W_{1}+\Omega',\ \ \ 
p_{\pm}=p_{1}\pm\ii   p_{2}
 \rightarrow exp(\pm\ii\Omega)p_{\pm};        \label{18:2}
\ee
here dot and prime denote differentiation with respect  to  $t$  and 
$r$, respectively. 

We choose the null NP tetrad as 
\be
l^{\mu}=(e^{-a},e^{-b},0,0), \ \
n^{\mu}=\frac{1}{2}(e^{-a},-e^{-b},0,0), \ \
m^{\mu}=\frac{1}{\sqrt{2}R}(0,0,1,\frac{\ii}{sin\vartheta}), 
                                                        \label{20}
\ee
the  non-vanishing NP coefficients being
$$
\rho=-\frac{1}{R}(\dot{R}e^{-a}+R'e^{-b}), \ \
\mu=\frac{1}{2R}(\dot{R}e^{-a}-R'e^{-b}), \ \
\gamma=-\frac{1}{4}(\dot{b}e^{-a}-a'e^{-b}), 
$$
\be
\varepsilon=\frac{1}{2}(\dot{b}e^{-a}+a'e^{-b}), \ \
\alpha=-\beta=-\frac{ctg\vartheta}{2\sqrt{2}R}.        \label{21}
\ee
The tetrad components of the gauge field potential are
\be
eA_{l}=(e^{-a}W_{0}+e^{-b}W_{1})\hat{L}_{3}, \ 
eA_{n}=\frac{1}{2}(e^{-a}W_{0}-e^{-b}W_{1})\hat{L}_{3}, \ 
\ii eA_{m}=\frac{1}{\sqrt{2}R} (p_{+}-1)\hat{L}_{+}, 
                                                     \label{23}
\ee
where    $\hat{L}_{\pm}=\hat{L}_{1}\pm\ii    \hat{L}_{2}$.     Using 
Eqs.(\ref{21}),(\ref{23}),    the    Dirac    equation    (\ref{16}) 
takes the form 
$$
\hat{\cal{D}}_{+}\Phi_{0}+\frac{1}{\sqrt{2}R}
(\hat{\Lambda}_{-}+\frac{1}{2}ctg\vartheta +(p_{-}-1)\hat{L}_{-}) 
~\Phi_{1}=\ii\mm~\Xi_{1},
$$
$$
\hat{\cal{D}}_{-}\Phi_{1} 
+\frac{\sqrt{2}}{R}
(\hat{\Lambda}_{+}+\frac{1}{2}ctg\vartheta-(p_{+}-1)\hat{L}_{+})
~\Phi_{0}=-\ii\sqrt{2}M~\Xi_{0},
$$
$$
\hat{\cal{D}}_{+}\Xi_{0}+\frac{1}{\sqrt{2}R}
(\hat{\Lambda}_{+}+\frac{1}{2}ctg\vartheta -(p_{+}-1)\hat{L}_{+}) 
~\Xi_{1}=-\ii\mm~\Phi_{1},
$$
\be
\hat{\cal{D}}_{-}\Xi_{1} 
+\frac{\sqrt{2}}{R}
(\hat{\Lambda}_{-}+\frac{1}{2}ctg\vartheta+(p_{-}-1)\hat{L}_{-})
~\Xi_{0}=\ii\sqrt{2}M~\Phi_{0},                  \label{24}
\ee
where
$$
\hat{\cal{D}}_{\pm}=e^{-a}(\partial_{t}+\frac{\dot{b}}{2}+
\frac{\dot{R}}{R}-\ii W_{0}\hat{L}_{1})\pm
e^{-b}(\partial_{r}+\frac{a'}{2}+
\frac{R'}{R}-\ii W_{1}\hat{L}_{1}),
$$
\be
\hat{\Lambda}_{\pm}=
\partial_{\vartheta}\pm\frac{\ii}{sin\varphi}\partial_{\varphi}. 
                                                  \label{25}
\ee
Introduce a pair of isotopic spinors:
\be
\chi^{+}=\left( \begin{array}{c} 
cos\frac{\vartheta}{2}\ e^{-\ii \varphi/2} \\
\\
sin\frac{\vartheta}{2}\ e^{\ii \varphi/2}\end{array} \right) ,
\ \ \ \
\chi^{-}=\left( \begin{array}{c} 
-sin\frac{\vartheta}{2}\ e^{-\ii \varphi/2} \\
\\
\ \ cos\frac{\vartheta}{2}\ e^{\ii \varphi/2}\end{array} \right) , 
                                                  \label{25:1}
\ee
with the following properties 
$$
\hat{L}_{1}\chi^{\pm}=\pm\frac{1}{2}\chi^{\pm},\ \ \
\hat{L}_{\pm}\chi^{\mp}=\chi^{\pm},\ \ \
\hat{L}_{\pm}\chi^{\pm}=0,
$$
\be
\hat{\Lambda}_{\pm}\chi^{\pm}=\frac{1}{2}ctg\vartheta\chi^{\pm},\ \ \
(\hat{\Lambda}_{\pm}+\frac{1}{2}ctg\vartheta)\chi^{\mp}=
\mp\chi^{\pm}.                                     \label{25:2}
\ee
To decouple angular variables in Eqs.(\ref{24}) we make the ansatz
$$
\Phi_{0}=\frac{1}{R}e^{-a/2}(\chi^{+}
F^{+}_{0}\ _{-1}Y_{jm}+\chi^{-} F^{-}_{0}\ _{0}Y_{jm}),\ \ \
\Phi_{1}=\frac{\sqrt{2}}{R}e^{-a/2}(\chi^{+}
F^{+}_{1}\ _{0}Y_{jm}+\chi^{-} F^{-}_{1}\ _{1}Y_{jm}),
$$
\be
\Xi_{0}=\frac{1}{R}e^{-a/2}(\chi^{+}
Q^{+}_{0}\ _{0}Y_{jm}+\chi^{-} Q^{-}_{0}\ _{1}Y_{jm}),\ \ \
\Xi_{1}=\frac{\sqrt{2}}{R}e^{-a/2}(\chi^{+}
Q^{+}_{1}\ _{-1}Y_{jm}+\chi^{-} Q^{-}_{1}\ _{0}Y_{jm}),
                                                  \label{26}
\ee
where $_{s}Y_{jm}(\vartheta,\varphi)$ are  spin  weighted  spherical 
harmonics  \cite{gold},  and   the   functions   $F^{\pm}_{\sigma}$, 
$Q^{\pm}_{\sigma}$ $(\sigma=0,1)$ depend on $t$  and  $r$  alone.  A 
gauge   transformation   of   the   form   (\ref{14:1}),(\ref{18:1}) 
induces the replacement 
\be
F^{\pm}_{\sigma}\rightarrow exp(\pm\ii\Omega/2)F^{\pm}_{\sigma},
\ \ \ 
Q^{\pm}_{\sigma}\rightarrow exp(\pm\ii\Omega/2)Q^{\pm}_{\sigma}. 
                                                      \label{26:1}
\ee
Taking into account Eqs.(\ref{25:2}) and also the following standard 
relations for the functions $_{s}Y_{jm}$ \cite{gold}
$$
(\partial_{\vartheta}\mp\frac{\ii}{sin\vartheta}
\partial_{\varphi}\pm s\ ctg\vartheta)\ _{s}Y_{jm}=
\pm\sqrt{(j\pm s)(j\mp s+1)}\ _{s\mp 1}Y_{jm},  
$$
one can  see  that  the  ansatz  (\ref{26})  solves  Eqs.(\ref{24}), 
provided that the following equations hold 
$$
\hat{D}_{+}^{\ast}F_{0}^{+}+\frac{\Lambda}{R}F^{+}_{1}=\ii M 
Q^{+}_{1},
$$
$$
\hat{D}_{-}^{}F_{1}^{-}-\frac{\Lambda}{R}F^{-}_{0}=
-\ii M Q^{-}_{0}, 
$$
$$
\hat{D}_{+}^{}F_{0}^{-}+\frac{p_{-}}{R}F^{+}_{1}+
\frac{\Lambda}{R}F^{-}_{1}=\ii M Q^{-}_{1}, 
$$
$$
\hat{D}_{-}^{\ast}F_{1}^{+}-\frac{p_{+}}{R}F^{-}_{0}
-\frac{\Lambda}{R}F^{+}_{0}=-\ii M Q^{+}_{0}, 
$$
$$
\hat{D}_{+}^{\ast}Q_{0}^{+}
-\frac{p_{+}}{R}Q^{-}_{1}
-\frac{\Lambda}{R}Q^{+}_{1}
=-\ii M F^{+}_{1},
$$
$$
\hat{D}_{-}^{}Q_{1}^{-}
+\frac{p_{-}}{R}Q^{+}_{0}
+\frac{\Lambda}{R}Q^{-}_{0}
=\ii M F^{-}_{0},
$$
$$
\hat{D}_{+}^{}Q_{0}^{-}-\frac{\Lambda}{R}Q^{-}_{1}
=-\ii M F^{-}_{1}, 
$$
\be
\hat{D}_{-}^{\ast}Q_{1}^{+}+\frac{\Lambda}{R}Q^{+}_{0}
=\ii M F^{+}_{0},                            \label{27}
\ee
where $\Lambda=\sqrt{j(j+1)}$, and 
$$
\hat{D}_{\pm}=e^{-a}(\partial_{t}+\frac{\dot{b}-\dot{a}}{2}+
\frac{\ii}{2}W_{0})\pm
e^{-b}(\partial_{r}+\frac{\ii}{2}W_{1}). 
$$
For the $S$-modes, $j=0$, the spherical harmonics  $_{\pm  1}Y_{jm}$ 
vanish,   and   the    resulting    equations    reduce    to    the 
$F^{+}_{0}=F^{-}_{1}=Q^{+}_{1}=Q_{0}^{-}=\Lambda=0$  case.  One  can 
see  that  the   transformations   (\ref{18:2}),(\ref{26:1})   leave 
Eqs.(\ref{27}) invariant. 

\section{Fermion energy spectrum}

For the background field quantities entering the system  (\ref{27}), 
we choose now (\ref{8}),(\ref{9}) and (\ref{13:1}),  describing  EYM 
fields of a vacuum-to-vacuum path. We treat  these  fields,  in  the 
adiabatic approximation, as a  sequence  of  instantaneously  static 
configurations, with $\lambda$ being a  parameter.  Our  aim  is  to 
study the energy spectrum of fermions as a function of $\lambda$.

In the chiral limit, the two chiralities decouple. In the following, 
we  consider  just  the  right-handed  spinor  component.   We   are 
interested in the $S$-wave sector dynamics, because, as  we  expect, 
it is in this sector where the level crossing phenomenon will occur. 
The following combinations are introduced for the fermions functions 
surviving in the case:
\be
F^{-}_{0}+F^{+}_{1}=\ii\sqrt{2}e^{-\ii \omega t}\psi^{+}(r),\ \ 
F^{-}_{0}-F^{+}_{1}=\sqrt{2}e^{-\ii \omega t}\psi^{-}(r).  \label{28}
\ee
The non-trivial equations (\ref{27}) take now the form 
$$
\left(\frac{d}{dr}+\frac{p_{1}}{N}\right)\psi^{+}+
\left(-\frac{\omega r^{2}}{\sigma N^{2}}
\left( 1+\frac{r^{2}}{4R_{\ast}^{2}}\right)^{-1}
+\frac{1}{2}W_{1}+\frac{p_{2}}{N}\right)
\psi^{-}=0,
$$
\be
\left(\frac{d}{dr}-\frac{p_{1}}{N}\right)\psi^{-}+
\left(\frac{\omega r^{2}}{\sigma N^{2}}
\left( 1+\frac{r^{2}}{4R_{\ast}^{2}}\right)^{-1}
-\frac{1}{2}W_{1}+\frac{p_{2}}{N}\right)
\psi^{+}=0,                                   \label{29}
\ee
where $N=\sqrt{r^{2}-2mr}$; the functions $m$ and $\sigma$ are given 
by (\ref{8}). The quantities specifying the background  gauge  field 
(\ref{9}) (taken in the regular gauge $II~$) read 
\be
W_{1}=\lambda w_{n}',\ \ 
p_{1}=cos\lambda cos w_{n}\lambda +w_{n}sin\lambda sin w_{n}\lambda, 
\ \ 
p_{2}=cos\lambda sin w_{n}\lambda -w_{n}sin\lambda cos w_{n}\lambda.
                                               \label{30}
\ee
Due to  (\ref{18:2}),  (\ref{26:1}),  these  equations  possess  the 
following gauge symmetry
$$
\psi^{+}\rightarrow \psi^{+}cos\frac{\Omega}{2} -
\psi^{-}sin\frac{\Omega}{2} , \ \ \
\psi^{-}\rightarrow\psi^{+} sin\frac{\Omega}{2} +
\psi^{-}cos\frac{\Omega}{2} ,
$$
\be
W_{1}\rightarrow W_{1}+\Omega',\ \ \
p_{1}\rightarrow p_{1}\ cos\Omega-p_{2}\ sin\Omega,\ \ \
p_{2}\rightarrow p_{1}\ sin\Omega+p_{2}\ cos\Omega,\    \label{31}
\ee
where $\Omega=\Omega(r)$. 
Sometimes it is convenient to utilize  this  symmetry  in  order  to 
search for the solutions  to  Eqs.(\ref{29}),  working  not  in  the 
regular gauge $II~$ for the background gauge field, but in the gauge 
$I~$ instead. The corresponding field parameters are simpler in this 
case: 
\be
W_{1}=0,\ \ \
p_{1}=cos^{2}\lambda+w_{n}sin^{2}\lambda,\ \ \
p_{2}=(1-w_{n}) sin\lambda cos\lambda~.                  \label{32}
\ee
Once the solution is found, one has to switch back  to  the  regular 
gauge $II~$ using the transformation (\ref{31}), with the  parameter 
\be
\Omega=\lambda (w_{n}-1).                             \label{32:1}
\ee

The conserved inner product (\ref{inn}) reads 
\be
<\psi_{2}|\psi_{1}>=\int dr\ 
\frac{r^{2}}{\sigma N^{2}}
\left( 1+\frac{r^{2}}{4R_{\ast}^{2}}\right)^{-1}
(\psi^{+}_{2}\psi^{+}_{1}+\psi^{-}_{2}\psi^{-}_{1}). \label{33}
\ee

The formal series expansions around  $r=0$  and  $r=\infty$  give  the 
following boundary conditions  for  the  normalizable  solutions  to 
Eqs.(\ref{29}), (\ref{30})
\be
\psi^{+}=O(r^{2}),\ \ \
\psi^{-}=O(r),\ \ {\rm as}\ \ r\rightarrow 0,       \label{33:1}
\ee
\be
\psi^{+}=O(\frac{1}{r}),\ \ \
\psi^{-}=O\left(\frac{R_{\ast}^{2}}{r^{2}}\right),\ \ \
{\rm as}\ \ r\rightarrow\infty.                    \label{33:2}
\ee
For each value of $\lambda$,  the  solutions  with  the  asymptotics 
(\ref{33:1}) and (\ref{33:2}) have to be matched  by  adjusting  the 
value of $\omega$. As a result, one  obtains  the  energy  spectrum, 
$\omega(\lambda)$. 

Before passing to tackling the problem numerically, it is helpful to 
investigate some particular cases. Firstly, consider  the  beginning 
of the vacuum-to-vacuum path, $\lambda=0$. Then, both  in  the  $I~$ 
and $II~$ gauges, one has $p_{1}=1$, $p_{2}=W_{1}=0$, that  is,  the 
gauge  field  is  zero;  also  $m=0$,  $\sigma=1$.    Eqs.(\ref{29}) 
describe in the case free fermions on $R\times S^{3}$. Introducing a 
new    radial    coordinate    $x\in[0,\pi]$     by     $r=R_{\ast}~ 
\left(1+r^{2}/4R_{\ast}^{2}\right)sin~x$,    one    can    represent 
Eqs.(\ref{29}) as
\be 
\left(\frac{d}{dx}+\frac{1}{sin~x}\right)\psi^{+}-
\frac{\nu}{2}\psi^{-}=0,\ \ \ 
\left(\frac{d}{dx}-\frac{1}{sin~x}\right)\psi^{+}+
\frac{\nu}{2}\psi^{-}=0,                                \label{34}
\ee
where  $\nu=2\omega  R_{\ast}$.  The  general  solution   of   these 
equations reads 
\be
\psi=\left(\begin{array}{c}\psi^{+} \\  \\
\psi^{-}\end{array}\right)=
A\left(\begin{array}{c}
\nu    cos(\nu x/2+B)    -ctg(x/2)\ 
sin(\nu x/2+B) \\  \\ 
-\nu    sin(\nu x/2+B)    +tg(x/2)\ 
cos(\nu x/2+B)\end{array}\right),              \label{35}
\ee
with $A$ and $B$ being integration constants; $\nu\neq\pm  1$.  This 
solution vanishes at $x=0$ only if  $B=0$.  When  $\nu$  is  an  odd 
integer, the solution vanishes also at $x=\pi$. The values  $\nu=\pm 
1$ correspond to the degenerate case.  The  solution  then  includes 
power terms, and it can not be regular both at  $x=0$  and  $x=\pi$. 
One obtains therefore the following spectrum of fermions  (see  also 
\cite{GS}) 
\be
\omega(\lambda=0)=\frac{\nu}{2R_{\ast}},\ \ \
\nu=\pm 3,\ \pm 5\ldots .                              \label{36}
\ee

Consider  now  the  end  point   of   the   vacuum-to-vacuum   path, 
$\lambda=\pi$. To find the spectrum, we observe that, in  the  gauge 
$I~$, the gauge field parameters (\ref{32})  take  the  same  values 
both at $\lambda=0$ and $\lambda=\pi$. This means that the  solution 
(\ref{35}) can be used also when  $\lambda=\pi$,  but  only  in  the 
gauge $I~$. To pass to  the  gauge  $II~$,  one  just  performs  the 
transformation (\ref{32:1}), so that the spectrum (\ref{36}) remains 
the same. 

We conclude therefore that, as $\lambda$ varies from zero to  $\pi$, 
the spectrum (\ref{36}) maps into itself. Certainly, this  does  not 
mean that each energy level maps  into  itself.  The  level  number, 
$\nu$, may change. For instance, some of the levels may cross  zero, 
ending   up   with   an   opposite   sign   of   their   energy   as 
$\lambda\rightarrow\pi$.  Observe   that,   in   the   gauge   $I~$, 
Eqs.(\ref{29}) are manifestly invariant under 
\be
\omega\rightarrow -\omega,\ \ \
\lambda\rightarrow\pi -\lambda,\ \ \
\psi^{+}\rightarrow -\psi^{+},\ \ \ 
\psi^{-}\rightarrow \psi^{-}.                          \label{37}
\ee
This means that normalizable solution with zero  energy  may  appear 
only if $\lambda=\pi/2$,  and  one  of  the  functions  $\psi^{\pm}$ 
vanishes. In the  gauge  $I~$,  the  corresponding  solution  (first 
discovered in Ref.\cite{GS}) can be found from (\ref{29}): 
\be
\psi^{+}=0,\ \ \ \psi^{-}=exp\left(\int^{r}
dr\frac{p_{1}}{N} \right);                          \label{38}
\ee
to  pass  to  the  regular  gauge  $II~$,   one   uses   (\ref{31}), 
(\ref{32:1}). 

Thus, we can see that the spectrum maps  into  itself  as  $\lambda$ 
runs from zero to $\pi$, and also some of the levels cross  zero  at 
$\lambda=\pi/2$. To get an  idea  about  the  whole  level  crossing 
picture, we use the following simple  analytic  approximation  which 
gives qualitatively the right result. Let us put in  (\ref{29})  for 
any $\lambda$,  $m=0$, $\sigma=1$, that is, we neglect the change of 
the geometry along the vacuum-to-vacuum path. Also, we  replace  the 
function $w_{n}$ in (\ref{30}), (\ref{32}) by a  step  function  $w$ 
obeying the same boundary conditions:
\be
w_{n}(r)\rightarrow w(r)=1,\ \ r<r_{0};
\ \ \ w(r)=-1,\ \ \ r>r_{0}.              \label{39}
\ee
Notice that the field topology remains intact, as  the  Chern-Simons 
number still changes in  accordance  with  Eq.(\ref{12}),  for  this 
number depends only on the  boundary  conditions  for  $w_{n}$  (see 
Eq.(\ref{5})). The normalizable zero energy solution (\ref{38}) also 
exists in this case. 

To find the solution for the fermion wave function, we observe that, 
in the gauge $I$, the function $w$ describes a field which  vanishes 
when  $r<r_{0}$  and  is  a   pure   gauge   when   $r>r_{0}$   (see 
Eq.(\ref{7})). The  resulting  solution  is  therefore  obtained  by 
matching the free fermion modes in  the  domain  $r<r_{0}$  and  the 
gauge transformed free modes in the region $r>r_{0}$. Let us use the 
same  independent  variable  $x=x(r)$   as   in   (\ref{34}).   When 
$x<x_{0}=x_{0}(r_{0})$,   the   regular   solution   is   given   by 
Eq.(\ref{35}) with $B=0$ and $A=1$. To obtain the  solution  in  the 
region $x>x_{0}$, one takes the solution (\ref{35})  with  arbitrary 
$A$ and  $B$,  and  performs  the  gauge  rotation  (\ref{31})  with 
$\Omega=2\lambda$  (to  find  $\Omega$   one   compares   (\ref{7}), 
(\ref{39}) and (\ref{18:1})). Matching these solutions at  $x=x_{0}$ 
specifies the values of $A$ and  $B$.  Next,  the  solution  in  the 
region $x>x_{0}$ has to be rotated to the gauge $II~$  and  required 
to vanish at $x=\pi$, which gives  the  quantization  condition.  It 
suffices to demand for only one component to vanish, say $\psi^{+}$, 
as the other will then vanish automatically, due to (\ref{33:2}). In 
the limit $x_{0}\rightarrow 0$, the quantization condition takes the 
following simple form
\be
ctg\lambda=\frac{2}{3}\nu x_{0}(1+O(x_{0}))+
\frac{1}{4}x_{0}^{2}(\nu^{2}-1)
(1+O(x_{0}))~tg(\nu\frac{\pi}{2}).                      \label{40}
\ee
This    equation    gives     the     spectrum     $\omega(\lambda)= 
\nu(\lambda)/2R_{\ast}$.    When    $\lambda=0,\pi$,    $ctg\lambda$ 
diverges, which requires $\nu=\pm 3,\pm 5,\ldots$, in agreement with 
Eq.(\ref{36}). As $\lambda$ runs from zero  to  $\pi$,  $ctg\lambda$ 
decreases from $+\infty$ to $-\infty$, which  forces  $\nu(\lambda)$ 
to  vary  from  $\nu(0)\neq  3$  to  $\nu(\pi)=\nu(0)-2$,  and  from 
$\nu(0)=3$ to $\nu(\pi)=-3$. 

We arrive, therefore, at the following level  crossing  picture  for 
the right-handed chiral fermions. When the  CS  number  varies  from 
zero to one, each energy level shifts downwards replacing 
the preceding one with 
the lower energy. The lowest positive energy level crosses zero when 
$N_{CS}=1/2$ and becomes  eventually  the  highest  negative  energy 
level. If the initial state corresponds to the fermion vacuum,  then 
one (left-handed) antifermion appears in the final state. 

Finally,  we  turn  to  the  numerical  analysis.  We  consider  the 
vacuum-to-vacuum path (\ref{9}), (\ref{13:1})  passing  through  the 
(lifted)  ground  state  EYM  sphaleron  field  configuration.   The 
compactification radius is chosen to be $R_{\ast}=10^{3}$  which  is 
much larger than the sphaleron size (the latter is of the  order  of 
unity; see Fig.1).  Numerical  integration  of  the  Dirac  equation 
(\ref{29}) with the boundary conditions  (\ref{33:1}),  (\ref{33:2}) 
gives the spectrum. The result is depicted in Fig.3 and confirms the 
qualitative picture described above. For the paths  passing  through 
the higher ($n>1$) EYM sphalerons, the  level  crossing  picture  is 
qualitatively the same. 

\section{Conclusion}
Our  results show  explicitly  the  anomalous  non-conservation   of 
fermion number in the  transition  processes  mediated  by  the  EYM 
sphalerons. It was  argued  recently  that  such  processes  can  be 
significant  at  high   energies   due   to   interference   effects 
\cite{main}.  One  may  expect  then  an  efficient  fermion  number 
non-conservation  in  the  high  energy  limit.  To  determine  more 
precisely the production of fermions, by going beyond the  adiabatic 
approximation, requires  a  numerical  solution  of  the  full  time 
dependent problem \cite{cott}, \cite{ZS}. Our results may be of  use 
also for further investigation of  the  quantum  correction  to  the 
classical energy of the sphaleron barrier, arising from the  fermion 
see \cite{diak}, \cite{NJ}.

\section*{Acknowledgments}
I would like to thank Professor N.Straumann for a careful reading of 
the  manuscript,  and  also  Jutta  Kunz,  A.Losev  and  A.Wipf  for 
discussions. 

This work was supported by  the  Swiss  National  Science 
Foundation.

\vspace{5 cm}
\section*{Figure captions}
\vspace{10mm}

Fig.1. Behaviour of the gauge field function, $w_{1}(r)$, the metric 
functions,  $m_{1}(r)$,  $\sigma_{1}(r)$,  and  the  radial   energy 
density $r^{2}T^{0}_{0}$ for the ground state $(n=1)$ EYM  sphaleron 
solution. In the units used, the sphaleron has the typical  size  of 
the order of unity. 

\vspace{5 mm}\noindent
Fig.2. The potential barrier dividing neighbouring  topological  EYM 
vacua. The top of the barrier is the position of  the  ground  state 
EYM sphaleron. 

\vspace{5 mm}\noindent
Fig.3. The fermion energy spectrum along the path which interpolates 
between distinct topological EYM vacua passing through  the  ground 
state sphaleron field. 

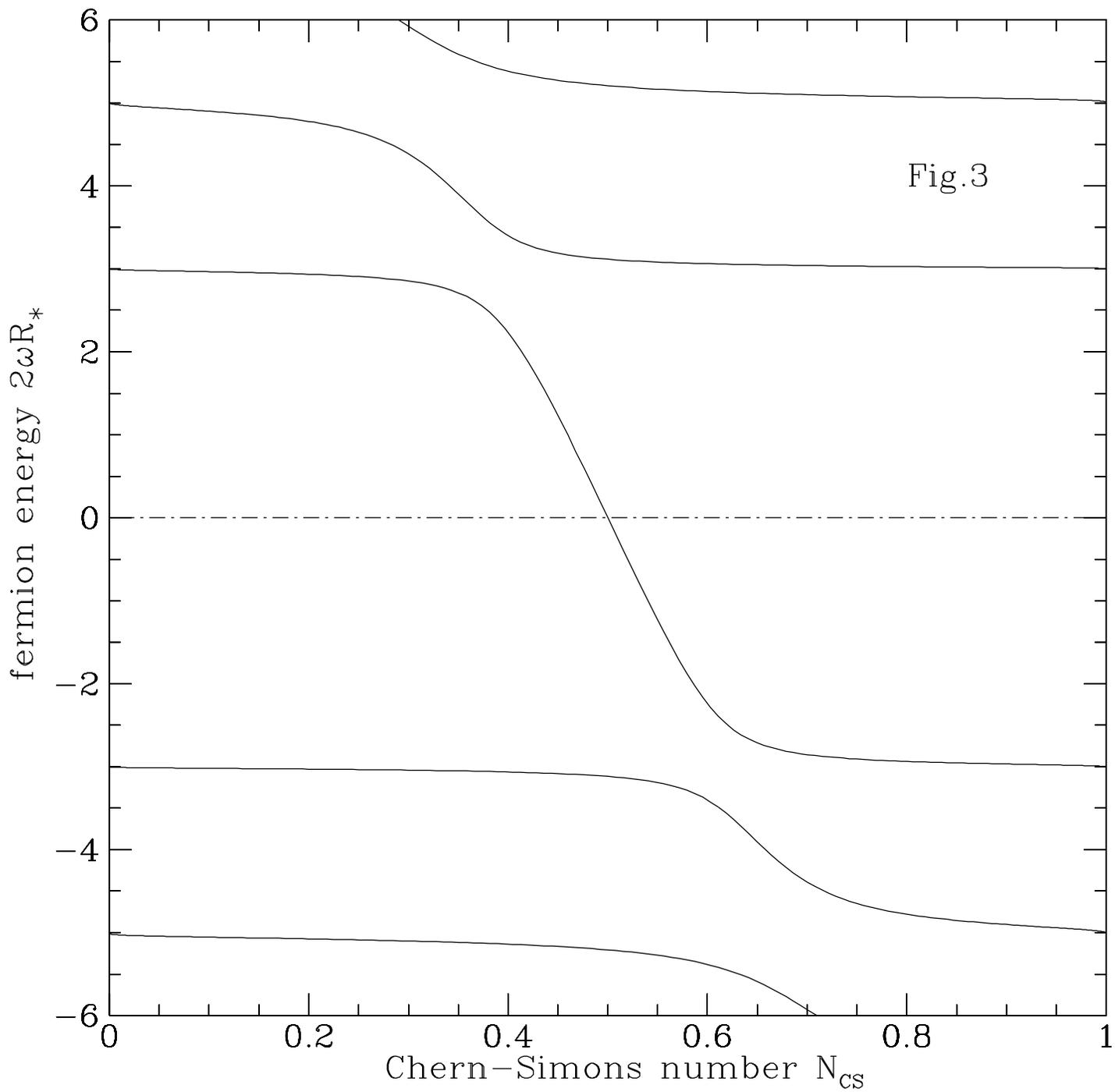
\end{document}